\documentclass[aps,prl,reprint,twocolumn,showpacs,superscriptaddress,amsmath,amssymb,citeautoscript]{revtex4-1}

\usepackage{graphicx}

\usepackage{color}

\usepackage{bm}% bold math
\renewcommand{\vec}[1]{\bm{#1}}

\begin{document}

\title{Ferromagnetic spin-orbital liquid of dipolar fermions in zigzag lattices}

\author{G. Sun}
\affiliation{Institut f\"ur Theoretische Physik, Leibniz Universit\"at Hannover, 30167~Hannover, Germany}

 \author{A. K. Kolezhuk}
\affiliation{Institute of High Technologies, Taras Shevchenko National University of Kiev,  03022 Kiev, Ukraine}
\affiliation{Institute of Magnetism, National Academy of Sciences and Ministry of Education,  03142 Kiev, Ukraine}

\author {L. Santos}
\author {T. Vekua}
\affiliation{Institut f\"ur Theoretische Physik, Leibniz Universit\"at Hannover, 30167~Hannover, Germany}

 \begin{abstract}
Two-component dipolar fermions in zigzag optical lattices allow for the
engineering of spin-orbital models.  We show that dipolar lattice fermions
permit the exploration of a regime typically unavailable in solid-state
compounds that is characterized by a novel spin-liquid phase with a finite
magnetization and spontaneously broken SU(2) symmetry. This peculiar spin
  liquid may be understood as a Luttinger liquid of composite particles
consisting of bound states of spin waves and orbital domain walls moving in an
unsaturated ferromagnetic background. In addition, we show that the system exhibits a 
boundary phase transitions involving non-local entanglement of edge spins.
\end{abstract}
\date{\today}
%
%67.85.Fg	Multicomponent condensates; spinor condensates
%67.85.Hj	Bose-Einstein condensates in optical potentials
%Spin-orbit coupling in condensed matter, 71.70.Ej
%phase transitions magnetic, 75.30.Kz
%quantum phase transitions, 05.30.Rt
%quantized spin models, 75.10.Jm
%quantum spin liquids, 75.10.Kt
%orbital, charge and other orders in, 75.25.Dk
%%%%%%%%%%%%%%%%%%%%%%%%%%%%%%%%%%%%%

\maketitle

% Introduction

\emph{Introduction.--} Frustrated spin systems provide a wealth of 
novel phenomena, both at the classical and quantum levels~\cite{Mila2011book}.
Frustration becomes particularly important in low-dimensional systems, 
where quantum and thermal fluctuations are strongly enhanced and
long-range order is suppressed. One of the most interesting
frustration-inducing mechanisms involves the interaction of spins with orbital
degrees of freedom~\cite{TokuraNagaosa00,Dagotto05,Mila2011book,Oles12rev}, which 
may result in spin-liquid states that lack 
long-range magnetic order~\cite{Feiner+97,KhaliullinMaekawa00,WangVishwanath09,Chaloupka+10,Corboz+11}.
However, in solid state systems controlling the strength of spin-orbital
interactions is hardly possible, limiting the exploration of spin-liquid phases.

Several recent works~\cite{Wu+07,Hermele+09,Gorshkov+10,Sun+12} have shown that ultra-cold
atomic gases in optical lattices can serve as quantum simulators of
spin-orbital models, providing the required freedom of controlling the effective
interactions by tweaking the optical lattice or by using Feshbach
resonances. Moreover, rapidly-developing experimental techniques make possible to study
the physics of higher energy bands, and to exploit orbital degeneracy~\cite{Wirth+10}. 

In particular, it has been recently shown~\cite{Sun+12} that spin-orbital models
of the Kugel-Khomskii type~\cite{KugelKhomskii82}, relevant in
transition metal oxides~\cite{TokuraNagaosa00,Dagotto05}, can be realized in
systems of dipolar spin-$\frac{1}{2}$ fermions loaded in doubly-degenerate
$p$-bands of optical zigzag lattices. For comparable on-site intra-orbital
repulsion $U$ and inter-orbital repulsion $V$, which is the typical situation in
solid-state scenarios~\cite{DiMatteo+04}, it was shown that dipolar fermions
have a rich ground state phase diagram containing states with
ferromagnetic (FM), antiferromagnetic (AF), dimerized and quadrumerized spin
order \cite{Sun+12}. Spin liquid phases are however absent in this regime.

Interestingly, contrary to the usual case in solid-state systems, a large ratio
$U/V$ may be attained for the case of dipolar fermions in zig-zag lattices by
properly controlling the ratio between dipolar and contact interactions. In this
Letter we show that for $U/V>2$ the ground state diagram contains a novel
spin-orbital liquid phase with a finite magnetization. This phase has a
spontaneously broken SU(2) spin symmetry and algebraically decaying longitudinal
spin correlations, while the orbital correlations decay exponentially. The
mechanism driving the transition into this phase is given by the softening of
composite excitations formed by bound states of spin waves and orbital domain
walls. We support our analytical arguments by numerical results
obtained by means of the density matrix renormalization group~(DMRG)
technique~\cite{White92,Schollwoeck05}. In addition, for open boundary conditions
we observe peculiar  boundary phase transitions that involve the formation of edge spins
that decouple from the bulk and get non-locally entangled.

%%%%%%%%%%%%%%%%%%%%%%%%%%%%
%% FIGURE 1
\begin{figure}[t]
\includegraphics[width=\columnwidth]{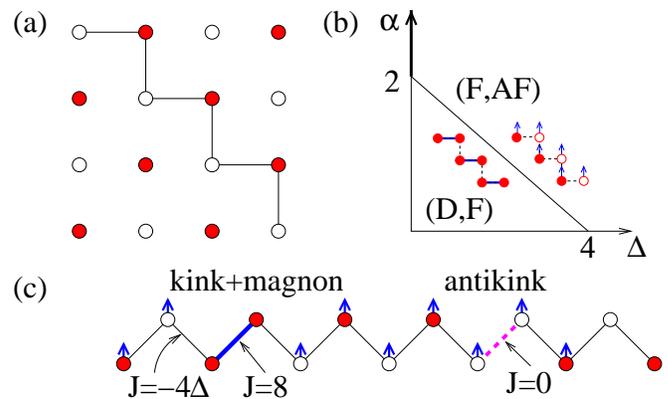}
\caption{ (Color online). (a) The path through a square lattice that
  defines the 1D zigzag lattice of spin-orbital model (\ref{ham-2d}); (b) phase
  diagram of the model for $\lambda=0$; (c) sketch of the kink-magnon bound state in the
  (F,AF) phase, where $J$ denotes the effective spin exchange
   on the corresponding link. A magnon binds only to the orbital
  kink, but not to the antikink.}
\label{fig:pathkink}
\end{figure}
%%%%%%%%%%%%%%%%%%%%%%%%%%%%

\emph{Spin-orbital model.--}  We consider two-component
(pseudo-spin-$1/2$) fermions loaded in doubly degenerate $p$-bands of a zig-zag
lattice (see Fig.\ \ref{fig:pathkink}(a), details on the experimental implementation of this system can be found
in Ref.~\cite{Sun+12}). The energy scales that determine the system are the
nearest-neighbor~(NN) hopping $t$ between equal orbitals, the average on-site
repulsion energies $U$~($V$) between same (different) orbitals, the Hund
coupling $J_{H}$, and an in-plane deformation of the optical lattice, distorting
the XY rotational symmetry of a single-site potential that mixes the orbitals
within the same site with an amplitude $\gamma$. In the Mott insulator
regime~(one fermion per site and strong coupling $U,V\pm J_H \gg t, \gamma $),
the system is described by an effective spin-orbital Hamiltonian (for details of
the derivation we refer to Ref.~\cite{Sun+12}):
\begin{eqnarray}
\label{ham-2d}
\mathcal{H}\!\!&=&\!\!\sum_{  l} (2\vec{S}_{l}\!\cdot\!\vec{S}_{l+1} \!+\!
\alpha- \frac{1}{2} ) (1+(-1)^l\sigma_{l}^{z}) (1+(-1)^l\sigma_{l+1}^{z})
\nonumber\\ 
&-& \Delta
\sum_{l} 2\vec{S}_{l}\cdot\vec{S}_{l+1}(1- \sigma_{l}^{z} \sigma_{l+1}^{z}) 
-\lambda \sum_{l} \sigma_{l}^{x}
\end{eqnarray}
where $\vec{S}_{l}$ are spin-$\frac{1}{2}$ operators acting on the lattice site
$l$, and $\sigma^{z,x}_{l}$ are the Pauli matrices describing the orbitals. The
parameters of the model, in the leading order in $J_{H}/U$, are given by
$\alpha\approx {U}/{V}$, $\Delta\approx {J_{H}U}/{V^{2}}$, $\lambda\approx
{\gamma U}/{t^{2}}$, and the Hamiltonian~(\ref{ham-2d}) has overall units of
$t^{2}/2U$. The dipole-cipole coupling is crucial because for
purely contact interaction there is no repulsion between fermions in the
spin-triplet state, and hence the Mott phase with one particle per site could
not stabilize.

\emph{Analytical estimates.--} Figure~\ref{fig:pathkink}(b) shows the phase
diaram for the case $\lambda=0$, for which the orbitals are classical. The
$\alpha>2$ region is dominated by a phase with FM spin order and AF orbital
order, which we label (F,AF) following the notation of Ref.~\cite{Sun+12}.
A smaller $\alpha$ favors the spontaneously dimerized (D,F) state,  with
the spin sector described by a product of singlets on even (odd) NN bonds and
ferromagnetic orbitals. On the line $\Delta=0$, $\lambda=0$, $\alpha>2$ the
spins are fully decoupled, whereas adding infinitesimally small
$\Delta$~($\lambda$) favors FM~(AF) spin exchange. This competition between
$\lambda$ and $\Delta$ leads to a first-order transition from the (F,AF) phase
to the (iH,AF) phase where the spin sector behaves as an isotropic Heisenberg
antiferromagnet, and the orbitals retain AF order. For $\lambda,\Delta\to 0$
this transition line can be easily estimated by computing the leading order
correction in $\lambda$ to the energy $E_{m}(k)$ of a magnon in the (F,AF)
state. For small momenta $k$, one obtains $E_{m}(k)\to
\big(2\Delta-\frac{\lambda^{2}}{8(\alpha-2)}\big)k^{2}$, collapsing at
$\lambda=\lambda_{F}= 4\sqrt{(\alpha-2)\Delta}+O(\Delta^{3/2})$.  A further
increase of $\lambda$ at fixed small $\Delta$ eventually leads to an Ising
transition in the orbital sector, bringing the system into the (iH,P) phase with
paramagnetic orbitals.

The (F,AF) ground state factorizes into a product of spin
and orbital wave functions, so there is a purely orbital Ising-type transition
from the (F,AF) phase to the (F,P) phase where orbitals are paramagnetic
(disordered) and spins remain fully polarized; the transition line thus can be
obtained exactly as $\lambda=\lambda_{\text{Ising}}=\alpha+\Delta/2$.

However, there is another, previously unknown, instability of the (F,AF) phase
which is of crucial interest here. This instability can be traced down to
the fact that in the (F,AF) phase magnons tend to bind to kinks in
the orbital order (see Fig.~\ref{fig:pathkink}(c)). If a kink-antikink
pair is excited on top of the (F,AF) state, on the link at the kink position the
effective exchange $J$ changes from ferromagnetic ($J\approx -4\Delta$ in 
zeroth order in $\lambda$) to antiferromagnetic ($J\approx 8$), acting as
an impurity which can bind a magnon. There is another impurity link with
$J\approx 0$ at the antikink position, but it does not support bound states.   

To the leading order in  $\lambda$, the energy of the kink-antikink
pair with a magnon bound to the kink is 
\begin{eqnarray} 
\label{kink} 
E_{\rm bs}(p,k)&=&4\alpha+2\Delta-8+8\Delta/(\Delta+4)\nonumber\\
&+& 2\lambda\big\{ 
[(8-\Delta^{2})/(4+\Delta)^{2}]\cos p-\cos k
\big\},
\end{eqnarray}
where $p$ and $k$ are the kink and antikink momenta, respectively. 
 The lower edge of this continuum is achieved at $p=\pi$, $k=0$, i.e., when the
magnon is essentially a propagating singlet dimer. This excitation softens at
$\lambda=\lambda_{c}= \frac{4}{3}(\alpha-2) +\frac{2}{9}(\alpha+4)\Delta + O(\Delta^{2})$. 
Hence, for $\lambda>\lambda_{c}$ a novel phase is expected with a finite density of
composite kink-dimer particles ``floating'' in the ferromagnetic
background~\cite{footnote}. An infinitesimal density of moving kinks and antikinks immediately
suppresses the orbital order, so the orbital AF order parameter experiences a
jump at the transition. Indeed, the (F,AF) product wave
function remains an exact eigenstate all the way up to
$\lambda=\lambda_{c}$, and $\lambda_{c}$ remains smaller than the
Ising transition value $\lambda_{\text{Ising}}$ in a finite range of $\Delta$. 
The ferromagnetic order in spins is retained, but the magnetization is no
more fully saturated.

%%%%%%%%%%%%%%%%%%%%%%%%%%%%
% FIGURE 2
\begin{figure}[t]
\includegraphics[width=0.48\textwidth]{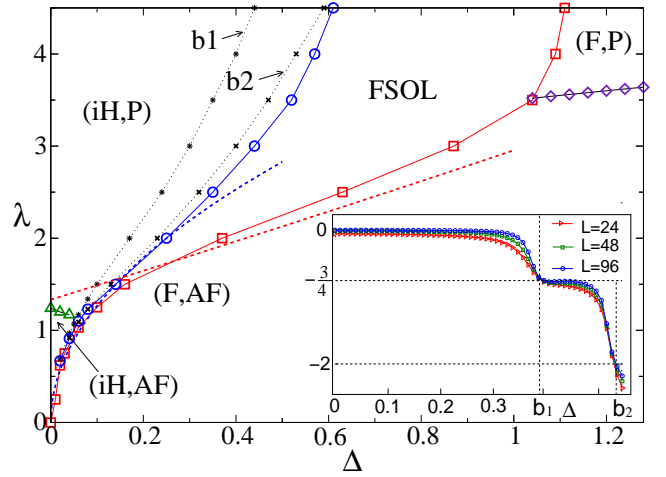}
\caption{ (Color online). Phase diagram of the 1D spin-orbital model for
  $\alpha=3$. Symbols denote numerical results (solid and dotted lines are a guide to the eye),  
whereas dashed lines correspond to the analytical estimates $\lambda_F$ and $\lambda_c$.  
Curves b1 and b2 mark the boundary phase transitions that involve non-local entanglement between edge spins 
$\vec\tau_1 = \vec{S_1}+  \vec{S_2}+ \vec{S_3}$ and $\vec\tau_N = \vec{S_N}+  \vec{S_{N-1}}+ \vec{S_{N-2}}$. 
The inset shows the ground-state correlation between edge spins $\langle
\vec\tau_1\vec\tau_N\rangle$ as a function of $\Delta$ for $\lambda=4$.}
\label{fig:phd-alpha3}
\end{figure}
%%%%%%%%%%%%%%%%%%%%%%%%%%%%
 
In the novel phase mentioned above, the SU(2) symmetry in the spin sector remains spontaneously broken, exactly as
in the (F,AF) phase, but the ground state belongs to a degenerate multiplet with
some spin $S_{\rm tot}<N/2$, where $N$ is the total number of particles ($N=L$ at unit filling considered here). This
phase is expected to have \emph{two} branches of gapless excitations, one with a
quadratic dispersion at small momenta (``spin'' mode, ferromagnetic magnons),
and the other with a linear dispersion (``charge'' mode, sound waves in the
Luttinger liquid of kink-dimer particles). This resembles the situation found in
spin-$\frac{1}{2}$ Bose gas, where such a spin-charge separation has been found
both in the 1D~\cite{Fuchs+05,Batchelor+06,Zvonarev+07,Kleine+08} and 
2D~\cite{ChungBhattacherjee08} cases. Since the longitudinal spin correlator is
related to the kink-dimer density fluctuations, it must decay
algebraically on top of the long-range order. This highly unusual phase can be
called a ferromagnetic spin-orbital liquid~(FSOL)~\cite{footnote2}.

%%%%%%%%%%%%%%%%%%%%%%%%%%%%
% FIGURE 3
\begin{figure}[tb]
\includegraphics[width=\columnwidth]{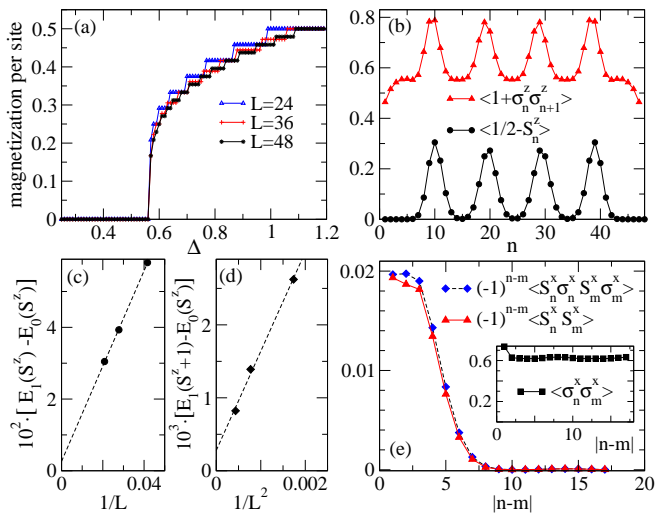}
\caption{(Color online). Properties of the FSOL phase for $\alpha=3$ and
  $\lambda=4$: (a) magnetization curve for different system sizes $L$; (b) magnon
  density~(circles) and orbital domain wall density~(triangles) in the
  ground state at $\Delta=0.86$, $S^{z}=20$, $L=48$; finite-size scaling of the particle-hole (c)
 and magnon (d) excitation gaps at $S^{z}=N/3$; (e)  ground
 state correlators for $\Delta=0.86$, $L=48$. See text for details.}
\label{fig:allferri}
\end{figure}
%%%%%%%%%%%%%%%%%%%%%%%%%%%%

\emph{Numerical results.--}
Our DMRG results confirm the analytical arguments given above. 
Figure~\ref{fig:phd-alpha3} shows our numerical results 
for the $(\lambda,\Delta)$ phase diagram of the 1D
version of the model~(\ref{ham-2d}) at $\alpha=3$. We considered open systems consisting of up to $L=96$ sites, 
monitoring different correlation functions, total spin of the ground state and fidelity susceptibility~\cite{Gu10rev} to detect phase boundaries. In addition, we
have checked our data on systems of up to $L=48$ sites with periodic boundary conditions. 
We have typically kept up to $800$ states (within a subspace with fixed
$S^{z}$) in our DMRG calculations.

We indeed observe the FSOL phase in a wide region of $\Delta$ and $\lambda$. As shown in Fig.~\ref{fig:allferri}(a), spontaneous magnetization
in the FSOL phase changes smoothly, confirming that there is no gap for single-particle excitations. 
In accordance with the composite-particle transition mechanism outlined
above, there is a clear correlation between the peaks in the densities of
orbital domain walls and magnons (see Fig.~\ref{fig:allferri}(b)). We have
checked that such a correlation persists at all magnetization values, and that the number of
peaks in the domain wall density is always equal to the number of magnons in the
ground state. 

Moreover, the energy of the lowest excitation in the same $S^{z}$ sector as the
ground state (the particle-hole gap) scales as $L^{-1}$ with the system size
$L$, while that in the sector corresponding to adding a magnon scales as
$L^{-2}$, as shown in Figs.~\ref{fig:allferri}(c) and (d). Thus these are two 
gapless excitation branches with linear and quadratic dispersion, respectively.  

Finally, in the FSOL phase the correlators $\langle
S^{x}_{l} S^{x}_{l'}\rangle$ and $\langle S^{x}_{l}\sigma^{x}_{l}
S^{x}_{l'}\sigma^{x}_{l'} \rangle$ are very close to each other, despite the
fact that $\langle \sigma^{x}_{l} \sigma^{x}_{l'}\rangle$ can be significantly
lower than one, see Fig.\ \ref{fig:allferri}(e) (in fact, for the
parameters presented in Fig.\ \ref{fig:allferri}(e) $\langle
S^{x}_{l}\sigma^{x}_{l} S^{x}_{l'}\sigma^{x}_{l'} \rangle$ is larger in
absolute value than $\langle S^{x}_{l} S^{x}_{l'}\rangle$ even though $|\langle
\sigma^{x}_{l} \sigma^{x}_{l'}\rangle| \simeq 6$ ).
 One can straightforwardly check that this follows from the fact that the wave
function of the bound state is very close to a singlet bond across the orbital
domain wall, so that the operator $S_{l}^{+}(1-\sigma_{l}^{x})$ nearly
anihilates the ground state. 

\emph{Boundary transitions.--} In addition to the existence of the FSOL phase,
the spin-orbital model with $\alpha>2$ is characterized by the appearance of
peculiar boundary phase transitions within the (iH,P) phase (curves b1 and b2 in
Fig.~\ref{fig:phd-alpha3}) at which the behavior of the edge spins in open
chains changes drastically.  When increasing $\Delta$ at fixed $\lambda$,
localized and strongly correlated $S=\frac{1}{2}$ edge spins emerge when
crossing the b1 curve. Further increasing $\Delta$ leads to a second transition
at the b2 line, where the value of the boundary spin changes from
$S=\frac{1}{2}$ to $S=1$. This effect is illustrated in the inset of
Fig.~\ref{fig:phd-alpha3}, where the correlation between edge spins is depicted
as a function of $\Delta$ for $\lambda=4$~\cite{footnote3}. We refer to the
Supplemental Material \cite{suppl} for further details.

\emph{Realization.--} Dipolar spinor fermions may be realized using polar
molecules (see Ref.~\cite{Sun+12} for a detailed discussion) or employing atoms
with large magnetic dipole moments, such as chromium~\cite{Lahaye+07}, dysprosium
~\cite{Lu+11}, or erbium~\cite{Frisch+13}.  For the particular case of
$^{53}$Cr, in a strong magnetic field, any two of the four lowest energy states
$|F,m_{F}\rangle=|\frac{9}{2},-\frac{9}{2}\rangle$, $|\frac{9}{2},
-\frac{7}{2}\rangle$, $|\frac{9}{2},- \frac{5}{2}\rangle$, and $|\frac{9}{2},
-\frac{3}{2}\rangle$, can be chosen to simulate the $\uparrow$ and $\downarrow$
pseudospin-$\frac{1}{2}$ states. Those states have approximately the same large
magnetic moments, given by the electronic spin projection $m_s=-3$, differing
only by their nuclear moment. The total interparticle potential is of the form
$ V(\vec{r}_{1}-\vec{r}_{2}) =\frac{\mu_0\mu^{2}}{4\pi |\vec{r}_{1}-\vec{r}_{2}
  |^3 } +g\delta(\vec{r}_{1}-\vec{r}_{2}) $, where $g={4\pi a_s \hbar^2}/{m}$
characterizes the contact interactions, where $a_s$ is the $s$-wave scattering
length, $m$ is the atomic mass, $\mu_0$ is the vacuum permitivity and $\mu$ is
the magnetic dipole moment. The interaction, resulting from the electronic
degrees of freedom, is pseudospin-independent, providing the desired
SU(2) spin symmetry of the problem. For $^{53}$Cr the natural value of ratio $U/V$, where \cite{Sun+12}
\begin{equation} 
U(V)=\int d\vec{r}_{1}d\vec{r}_{2}\,
  {p^2_{\alpha}}(\vec{r}_{1}) \, V(\vec{r}_{1}-\vec{r}_{2}) \,
  {p^2_{\alpha(\beta\neq \alpha)}}(\vec{r}_{2}), 
\end{equation}
(here $\alpha, \beta=x,y$ and $p_{x,y}(\vec{r})$ are the orbital wave functions
centered at the same site) is in the regime $U/V > 2$ considered in this letter.

\emph{Summary.--} We have shown that dipolar two-component fermions loaded in
the $p$-bands of a zigzag optical lattice may allow for the realization of a
novel,  unaccessible in solid-state systems, spin-orbital liquid phase
characterized by a finite but unsaturated magnetization. This phase, like ferromagnet, has spontaneously broken SU(2) symmetry, but, unlike a
ferromagnet, it has algebraically decaying longitudinal spin correlations.  This
phase can be viewed as a Luttinger liquid of bound composites of singlet spin
dimers and orbital domain walls on top of a fully polarized ferromagnetic phase.

%\acknowledgements
\emph{Acknowledgments.--} We thank  H. Frahm, T. Osborne, and H.-J. Mikeska for helpful discussions. 
This work has been supported by QUEST (Center for
Quantum Engineering and Space-Time Research) and DFG Research Training Group
(Graduiertenkolleg) 1729.

\section{Supplementary materials on:  Details of boundary transitions in the (\lowercase{i}H,P) phase}

%%%
\begin{figure}[t]
\includegraphics[width=\columnwidth]{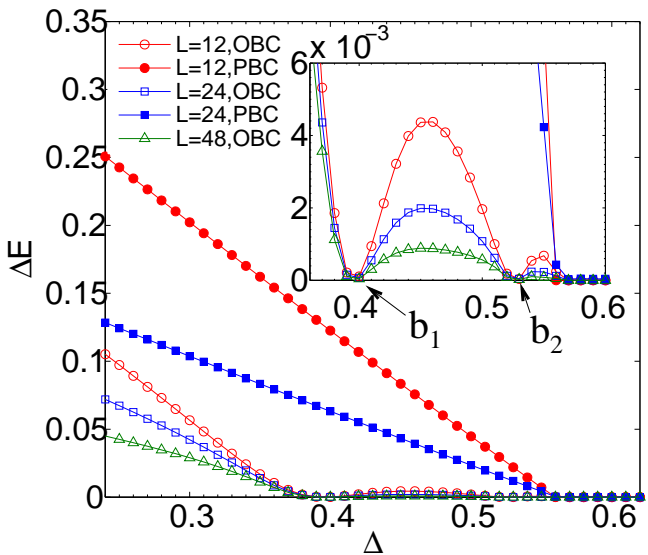}
\caption{(Color online). Excitation gap for open boundary conditions~(OBC) and periodic boundary conditions~(PBC) as a function of $\Delta$. 
The inset zooms in the region $0.36<\Delta<0.6$, revealing the existence of the two boundary transtions at b1 and b2.}
\label{fig:Gaps}
\end{figure}
%%%

\begin{figure}[t]
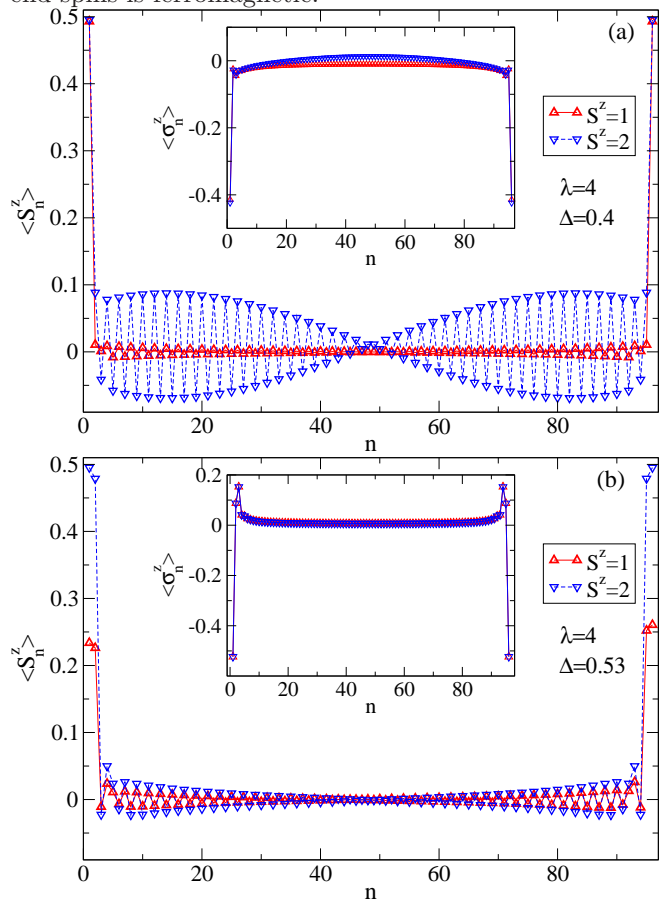

\includegraphics[width=\columnwidth]{EdgeS-SIG-M1.eps}
\includegraphics[width=\columnwidth]{EdgeS-SIG-M2.eps}
\caption{(Color online). Spin and orbital distribution profiles in the ground
  states of the $S^{z}=1$ and $S^{z}=2$ sectors, at two points of the $\alpha=3$
  phase diagram, along the $\lambda=4$ line: (a) at a point between the b1 and
  b2 boundary transition lines, the $S^{z}=1$ excitation is localized at the
  system edges, while the $S^{z}=2$ excitation belongs to the bulk; (b) at
  another point between the b2 line and the boundary of the FSOL phase, both
  $S^{z}=1$ and $S^{z}=2$ excitations are localized at the edges.}
\label{fig:edgespins}
\end{figure}

The boundary transitions inside the (iH,P) phase are peculiar in 1D, since edge spins are 
separated by a macroscopic distance, and the only way to communicate between them is through the bulk from which 
they effectively decouple. To prove that we are dealing 
with a boundary phenomenon we compare the excitation gaps for open and periodic boundary conditions. 
One can clearly see from Fig.~\ref{fig:Gaps} that low-lying states below the
bulk modes develop for open boundary conditions. 
Another illustration  of this boundary transition is provided by Fig.~\ref{fig:edgespins},
which shows the behavior of the first excited states in the $S^{z}=1$ and $S^{z}=2$ sector.

To describe the physics of this transition at the qualitative level, it is
instructive to consider the limit of large $\lambda$.
In the strong $\lambda$ limit one can integrate out
orbital degrees of freedom to obtain an effective spin-$\frac{1}{2}$ model. In
the leading order in $1/\lambda$, its Hamiltonian has the form of a
$J_{1}$-$J_{2}$ model with modified first and last nearest-neighbor links:
\begin{eqnarray}
\label{J1J2}
H_{S} &=& j_{1} \sum_{n=2}^{N-2}\vec{S}_{n}\cdot\vec{S}_{n+1} 
  +j_{2}\sum_{n=2}^{N-1} \vec{S}_{n-1}\cdot\vec{S}_{n+1}\nonumber\\
&+&
j_{1}'(\vec{S}_{1}\cdot\vec{S}_{2} + \vec{S}_{N-1}\cdot\vec{S}_{N}),
\end{eqnarray}
where 
\begin{eqnarray} 
\label{j1j2pars} 
&& j_{1}=2(1-\Delta)+\frac{4+(1+\Delta)(\Delta+2-2\alpha)}{2\lambda},\nonumber\\
&& j_{2}=1/\lambda,\quad 
j_{1}'=j_{1}+\frac{1-2\alpha}{2\lambda}.
\end{eqnarray}
One can see that with increasing $\Delta$, the boundary link strength $j_{1}'$
goes through zero at some point and changes its sign to a ferromagnetic
coupling. This effectively creates 'impurity' spins attached ferromagnetically
at the ends of the spin-$\frac{1}{2}$ chain. Interaction between the end spins
is mediated by the bulk. For an even number of sites the effective interaction
is antiferromagnetic, whereas for odd number of sites the effective interaction
between the end spins is ferromagnetic.

The second boundary transition, $b_2$, is similar in nature to the first one,
but now the last two spins decouple from the bulk creating an effective spin-$1$
localized at each boundary and ferromagnetically attached to the
anitferromagnetic spin-$\frac{1}{2}$ chain. The interaction between the spin-1
edge impurities is antiferromagnetic for even number of sites and ferromagnetic
for odd number of sites. The lowest excitations are boundary excitations: a
boundary triplet with total spin $S^T=1$ and a little higher boundary quintet
with spin $S^T=2$.

%%%
%%%


\begin{thebibliography}{10}
\bibitem{footnote} The expression for $\lambda_c$, obtained from the
perturbation theory in $\lambda$, is formally valid when $\Delta$ and $(\alpha-2)$ are both small.
With increasing $\lambda$, the first instability of the (F,AF) phase is
determined by $\lambda_F$ at very small $\Delta$, but starting from
$\Delta\simeq 0.14(\alpha-2)$ it is described by $\lambda_c$.
\end{thebibliography}
\end{document}